\documentclass[pra,twocolumn,a4paper,showpacs,floatfix]{revtex4}

\usepackage{graphicx} 
\usepackage{amsmath,amssymb}

\begin{document}

\title{Weak disorder expansion for localization lengths of quasi-1D
  systems}

\author{Rudolf A.\ R\"{o}mer}
\affiliation{Department of Physics and Centre for Scientific Computing,
  University of Warwick, Coventry CV4 7AL, United Kingdom}

\author{Hermann Schulz-Baldes}
\affiliation{Institut f\"{u}r Mathematik, Strasse des 17.\ Juni 136,
  Technische Universit{\"a}t Berlin, 10623 Berlin, Germany}


\begin{abstract}
  A perturbative formula for the lowest Lyapunov exponent of an Anderson
  model on a strip is presented. It is expressed in terms of an
  energy dependent doubly stochastic matrix, the size of which is
  proportional to the strip width.  This matrix and the resulting
  perturbative expression for the Lyapunov exponent are evaluated
  numerically. Dependence on energy, strip width and disorder strength
  are thoroughly compared with the results obtained by the standard
  transfer matrix method.  Good agreement is found for all energies in
  the band of the free operator and this even for quite large values of
  the disorder strength.
\end{abstract}


\maketitle

The Anderson model describes the generic behavior of the
motion of an electron in a disordered solid. In the one-dimensional (1D)
situation, rigorous proofs of strong localization have been given \cite{GMP}.
Moreover, the localization length has been calculated
\cite{Tho79,PasF92} for weak disorder and its inverse, the Lyapunov
exponent, is given by $\gamma\approx w^2/(96 \sin^2 k) = w^2/(96
-24\,E^2)$, where $E=2 \cos k$ is an energy in the Bloch band of the
unperturbed operator (away from band center and edges) and $w$ is the
disorder strength in the normalization discussed below.  For the 2D
case, scaling theory \cite{AbrALR79} predicts also strong localization
with an disorder dependence of the localization length of the form
$\exp(1/w^2)$. This was confirmed by high-precision numerical studies
based on the transfer-matrix method (TMM) \cite{MacK81,PicS81}.  In 3D,
a transition to the so-called weak localization regime with diffusive
motion is expected for low disorder strength and energies in the band of
the free operator.  A rigorous proof of strong localization in 2D and 3D exits
only for the band edges and at high disorder \cite{FS,AM} and, in
particular, not for energies in the band and small disorder
in the 2D situation. No
rigorous breakthrough results are known for the weak localization
regime.

In order to approach the higher dimensional cases, a detailed
understanding of the quasi-1D situation, {\sl i.e.}  an infinite wire
with many channels, is of crucial importance.  The TMM
\cite{MacK81,PicS81} is here a very reliable tool for a numerical study
of the inverse localization length, namely the smallest Lyapunov
exponent.  Thouless argued that it is proportional to $w^2/L$ where $L$
is the number of channels \cite{Tho77}. The Dorokhov-Mello-Pereyra-Kumar
theory gives a microscopic derivation of this behavior (see \cite{Ben97}
for a review), but is not based on a calculation starting directly from
an original model and hence does not allow to study finer properties
such as the energy dependence of the localization length. A perturbative
analytical calculation of the smallest Lyapunov exponent of an Anderson
model on a strip was recently given by one of the authors \cite{Sch03}.
The techniques of that work also allow to deal with other models having
symplectic transfer matrices.

In this letter, we first recall the perturbative formula from
\cite{Sch03} and then compare it numerically with the TMM. Our main
result is that the perturbative formula works remarkably well for all but a
discrete set of energies and, quite surprisingly, relatively large values
of the disorder strength. This allows to understand the rich structure
of the energy dependence of the smallest Lyapunov exponent. We also
study various quantities associated to the random dynamical system
underlying the TMM.

Let us begin by recalling that the Anderson Hamiltonian
on an infinite strip of finite width $L$ is given by
\begin{equation}
\label{eq-Hand}
{\bf H}
\;=\;
- \sum_{ \langle{\bf x}, {\bf y}\rangle } | {\bf x} \rangle\langle {\bf y} |
  + w \sum_{{\bf x}} v({\bf x}) | {\bf x} \rangle\langle {\bf x} |
\;,
\end{equation}
with tight-binding states $| {\bf x} \rangle$ at ${\bf x}=(n,m)$
where $n\in{\mathbb Z}$ and $m=1,\ldots, L$. The hopping is
between nearest neighbors $ \langle{\bf x}, {\bf y}\rangle$ only
and with periodic boundary conditions in the $m$-direction. The
$v({\bf x}) \in [-\frac{1}{2}, \frac{1}{2}]$ are independent and
identically distributed random variables with (for sake of
concreteness \cite{choice}) uniform distribution so that ${\bf
E}\left[v({\bf x})^2\right]=\frac{1}{12}$. For simplicity, let us
also choose $L$ even. The Schr\"odinger equation ${\bf
H}\Psi=E\Psi$ is rewritten in a recursive form using the $2L\times
2L$ transfer matrices
\begin{equation}
\label{eq-tn}
{\bf T}(n)
\;=\;
  {\left(
      \begin{array}{cc}
        {\bf \Delta}_L + w {\bf V}(n) - E {\bf 1}& {\bf -1} \\
        {\bf 1} & {\bf 0}
      \end{array}
    \right)}
\;,
\end{equation}
where ${\bf \Delta}_L$ is the discrete Laplacian in the transverse
direction with periodic boundary conditions and ${\bf V}(n)={\rm
diag}(v(n,1), \ldots, v(n,L))$.
The transfer matrix is symplectic, namely, ${\bf T}(n)^{t} {\bf J}
{\bf T}(n)= {\bf J} = {\left(\begin{array}{cc}{\bf 0} & {\bf -1} \\
{\bf 1} & {\bf 0}\end{array} \right)}$.
Associated with this family of random matrices are the Lyapunov
exponents $\gamma_1 \geq \ldots \geq \gamma_L \geq 0$ defined via
\begin{equation}
\label{eq-lyapdef}
\sum_{q=1}^{p} \gamma_q
\;=\;
\lim_{N\rightarrow\infty} \frac{1}{N}
\log \left( \left\| \prod_{n=1}^{N} \Lambda^{p} {\bf T}(n)
\right\| \right)
\;,
\end{equation}
for $p=1, \ldots, L$, where $\Lambda^p$ denotes the $p$-fold exterior
product. They are self-averaging quantities so that an average over the
disorder configurations may be taken before the large $N$ limit
\cite{BL}.

The first aim is to bring the transfer matrix at $w=0$ to its symplectic
normal form. Let $\mu_l= - 2 \cos\left( 2 \pi l / L\right) - E$, $l=0,
\ldots, L-1$, denote the eigenvalues of ${\bf \Delta}_L-E$. Clearly,
$\mu_l= \mu_{L-l}$ so that all eigenvalues $\mu_l$ except for $l=0, L/2$
are doubly degenerate. If $|\mu_l| < 2$, we call it an {\em elliptic}
eigenvalue and define its {\em rotation phase} $\eta_l$ by $\mu_l=
e^{\imath \eta_l} + e^{-\imath \eta_l}$. If on the other hand, $|\mu_l|
> 2$, we call it {\em hyperbolic} and define its {\em dilation exponent}
$\eta_l$ by $\mu_l= e^{\eta_l} + e^{- \eta_l}$.  There are
$\frac{L}{2}-1$ energies $E$ within the band $[-4,4]$ of the free strip
Laplacian for which the parabolic case $|\mu_l|=2$ occurs. These
energies corresponding to {\em interior band edges} are for now
excluded, but will be further discussed below.  Using the corresponding
eigenvectors of ${\bf \Delta}_L$ it is possible to construct a
symplectic matrix ${\bf M}$ such that
\begin{equation}
\label{eq-basischange}
{\bf M}^{-1} {\bf T}(n) {\bf M}
\;=\;
{\bf R} \left[{\bf 1} + w {\bf P}(n) \right]
\;.
\end{equation}
The symplectic matrix ${\bf R}$ is built from elliptic and
hyperbolic rotation matrices ${\bf R}_e(\eta)$, ${\bf R}_h(\eta)$,
respectively given by
\begin{equation}
\label{eq-rotmat}
    \left( \begin{array}{cc}
    \cos\eta & -\sin\eta \\
    \sin\eta & \cos\eta
    \end{array}\right),
\quad
    \left( \begin{array}{cc}
    \cosh\eta & \sinh\eta \\
    \sinh\eta & \cosh\eta
    \end{array}\right)
\;.
\end{equation}
More precisely, the 4 entries of ${\bf R}$ at $(l,l)$, $(l,L+l)$,
$(l+L,l)$ and $(l+L,l+L)$ form the matrices ${\bf R}_e(\eta_l)$, ${\bf
  R}_h(\eta_l)$ depending on whether $\mu_l$ is elliptic or hyperbolic.
All other entries of ${\bf R}$ vanish.
The matrix ${\bf P}(n)={\bf M}^{-1} \left( \begin{array}{cc}0 & 0\\{\bf
      V}(n) & 0
\end{array}\right) {\bf M}$
is nilpotent and in the Lie algebra of the symplectic group.

These free modes naturally group themselves into channels which
are the even-dimensional subspaces of ${\mathbb R}^{2L}$ rotating
under ${\bf R}$ with the same frequency. In our situation, there
are $L/2 +1$ such channels indexed by $l=0, \ldots, L/2$. The
$l$th channel is given by the components $l$, $L-l$, $L+l$, $2L-l$
of ${\mathbb
  R}^{2L}$ and we denote the corresponding projection by $\pi_l$. The
channels $l= 0, L/2$ are simple, while all others are doubly degenerate.

Let us introduce a symplectic frame $u= (u_1, \ldots, u_L)$ to be
a set of orthonormal vectors in ${\mathbb R}^{2L}$ satisfying
skew-orthogonality $\langle u_p | {\bf J} | u_q \rangle = 0$ for
all $p,q= 1, \ldots, L$.  Given an initial symplectic frame
$u(0)$, new frames $u(n)$ are constructed iteratively as follows:
apply ${\bf R} \left[ {\bf 1} +
  w {\bf P}(n) \right]$ to each vector of $u(n-1)$ and then use
Gram-Schmidt orthonormalization procedure in order to obtain $u(n)$.
This gives a random dynamical system on the space of symplectic frames
(which is isomorphic to the $L$-dimensional unitary group).  The
advantage of the basis change (\ref{eq-basischange}) is that the
discrete-time dynamics of frames at $w=0$ is simply given by rotations.
A weak random potential perturbs this simple dynamics in an analytically
controllable way.

Important in our perturbative formula will be the weight
$\rho_{p,k}(n) = \langle u_{p}(n) | \pi_k | u_{p}(n) \rangle$ of
the $p$th frame vector in the $k$th channel at iteration $n$, as
well as its Birkhoff mean
\begin{equation}
\label{eq-rhomean}
\langle \rho_{p,k} \rangle
\;=\;
\lim_{N\rightarrow\infty}\frac{1}{N}
\sum_{n=0}^{N-1} \rho_{p,k}(n)
\;.
\end{equation}
Each matrix $\rho_{p,k}(n)$ is doubly stochastic, namely, the sum
over the channel index $k=0, \ldots, L/2$ equals $1$ while the sum
over the frame vector index $p=1, \ldots, L$ is equal to the
degeneracy of the $k$th channel. The latter fact is related to the
symplectic structure of the frame and $[\pi_l,{\bf J}]=0$.  In a
similar fashion, one can define other Birkhoff averages such as
$\langle \rho_{L,l} \rho_{L,k} \rangle$.

\begin{figure}
  \centering
  \includegraphics[width=0.95\columnwidth]{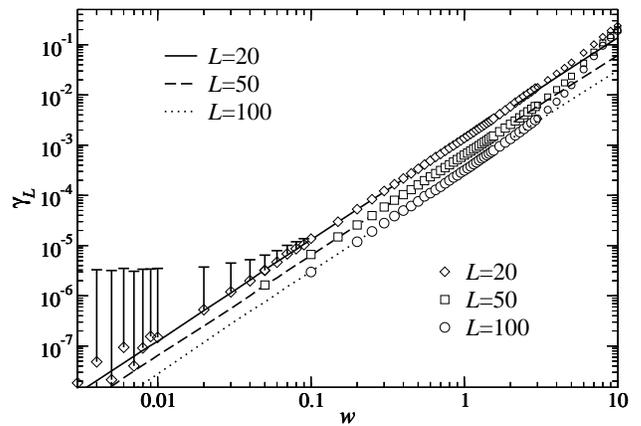}
  \caption{\label{fig-1}
    Lyapunov exponent $\gamma_L$ as function of disorder strength for energy
    $E=1.1$ and strip widths $L= 20, 50, 100$. Lines have been computed
    from (\ref{eq-lyapexpansympgeneral}) with $0.05\%$ error, symbols
    denote results of TMM calculations (with $0.5\%$ error if there is
    no error bar). Error bars are drawn on one side only.  }
\end{figure}

The rigorous perturbative formula for the smallest Lyapunov
exponent within the band is then, under a hypothesis on the
incommensurability of the rotation phases which excludes energies
with Kappus-Wegner-type anomalies \cite{KapW81} and interior band
edges, given by
\begin{equation}
\label{eq-lyapexpansympgeneral}
\gamma_L
\;=\;
\frac{w^2}{96 \,L}
\sum_{l,k} \frac{2-\delta_{l,k}}{\sin\eta_l \sin\eta_k} \;
\langle\rho_{L,l} \; \rho_{L,k}\rangle
\;+\;
{\cal O}(w^4)
\;,
\end{equation}
where the sum runs over elliptic channels only (actually, for hyperbolic
channels $k$, $\rho_{L,k}(n)$ almost vanishes for $n$ large enough as
will be discussed below). For a single channel, this expression reduces
to the perturbative 1D result given above if one sets ${\bf \Delta}_1=0$
in (\ref{eq-tn}) and then $\mu_0=E$ so that $\eta_0=k$.  We emphasize
that only the averaged channel weights of the last frame vector $u_{L}$
enter expression (\ref{eq-lyapexpansympgeneral}).  We also remark that
the dependence of the error term on $L$ and $E$ remains unspecified.

For a numerical study of (\ref{eq-lyapexpansympgeneral}), one
first evaluates the Birkhoff averages $\langle \rho_{L,l}
\rho_{L,k} \rangle$ as in (\ref{eq-rhomean}) by generating random
transfer matrices just as in the TMM \cite{MilRSU00}.  Reporting
this into (\ref{eq-lyapexpansympgeneral}) and neglecting the
${\cal O}(w^4)$ term gives the perturbative values plotted in
Fig.~\ref{fig-1} (as well as Fig.~\ref{fig-2} and \ref{fig-3}
below).  This was done for a typical energy $E=1.1$ away from
internal band edges such that $|\mu_l|\neq 2$ for all $l$. For
comparison, we also plot the value of $\gamma_L$ as evaluated by
the standard TMM. Very good agreement is obtained even for rather
large values of $w$. For very small $w$, the numerical convergence
of the TMM estimates for $\gamma_L$ is computationally intensive
whereas the convergence of the average channel weights needs about
a factor $10^3$ less iterations. More stable results can hence be
obtained at a fraction of the computational cost. For $L=20$ and
$E=1.1$, the validity of the perturbative formula breaks down at
about $w\approx 3.0$, for large $L$ a bit earlier. The breakdown
happens in the region of crossover from quasi-1D to 2D behavior,
i.e. $L\gamma_L\approx 1$.

\begin{figure}
  \centering
  \includegraphics[width=0.95\columnwidth]{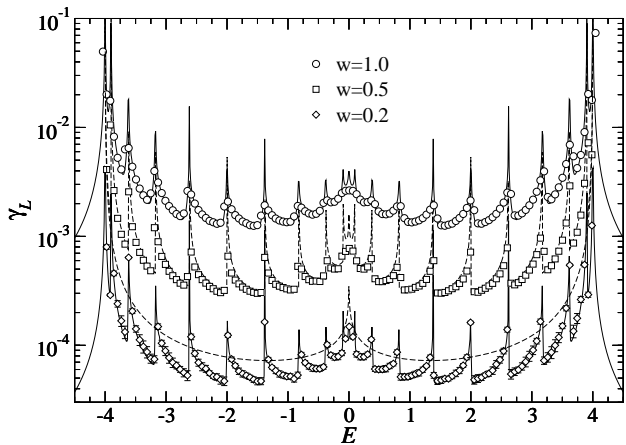}
  \caption{\label{fig-2}
    Lyapunov exponent $\gamma_L$ as function of energy for disorder
    $w=0.2, 0.5, 1.0$ and strip width $L=20$. Solid lines have been
    computed from (\ref{eq-lyapexpansympgeneral}) with $0.05\%$ error,
    symbols denote results of TMM calculations with $0.5\%$ error. For
    $w=0.5$ the dashed solid line was calculated from Eq.\
    (\ref{eq-approx}) and for $w=0.2$ from Eq.\ (\ref{eq-envelop}).}
\end{figure}

\begin{figure}
  \centering
  \includegraphics[width=0.95\columnwidth]{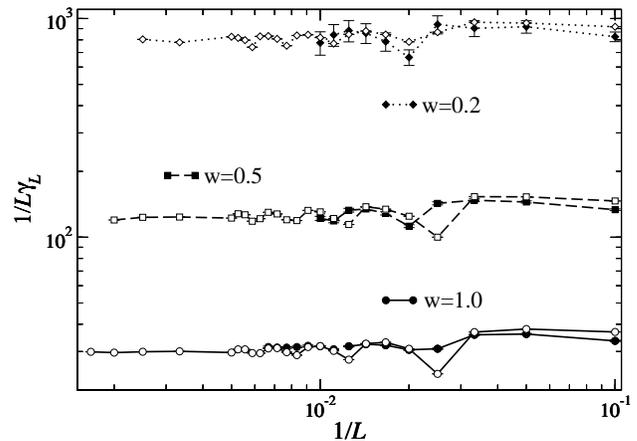}
  \caption{\label{fig-3}
    Log-log plot of the reduced localization length $1/L \gamma_L$ as
    function of inverse strip width $1/L$ for energy $E=1.1$ and
    disorders $w=0.2$ ($\diamond$), $w=0.5$ ($\Box$), and $w=1.0$
    ($\circ$). Full symbols denote results of TMM calculations with
    $0.5\%$ error, open symbols have been computed using
    (\ref{eq-lyapexpansympgeneral}) with $0.05\%$ error. The lines serve
    as guide to the eye. }
\end{figure}

Fig.~\ref{fig-2} shows the energy dependence of $\gamma_L$ and its
perturbative approximation (\ref{eq-lyapexpansympgeneral}). The
oscillatory behavior is very well reproduced by the internal band
edges corresponding to some eigenmode of ${\bf \Delta}_L$, that is
the peaks lie precisely on energies $E$ where $|\mu_l|=2$ for some
$l$ so that ${\sin\eta_l}=0$. As the validity of
(\ref{eq-lyapexpansympgeneral}) in their vicinity is restricted,
the singularities of the perturbative approximation are
artificial. In fact, one has to adapt the analysis of \cite{DG84}
using the normal form of \cite{Sch04b} to this higher dimensional
case. This will be done elsewhere.

%
%
%

Fig.~\ref{fig-3} shows the strip width dependence of the reduced
localization length in a plot usually used to infer the universal 2D
scaling function. The scaling hypothesis \cite{AbrALR79,KraM93} states
that there is a unique scaling function $F$ such that one can find a
function $\xi(w)$, also called the 2D localization length, so that all
the data $\gamma_L(w)$ satisfies $\log\left(\frac{1}{L\gamma_L(w)}\right)
=F\left[\log\left(\frac{\xi(w)}{L}\right)\right]$.
Hence Fig.~\ref{fig-3} allows to read off a part of $F$ which is usually
difficult to determine numerically
because the TMM converges badly for very small
$w$. Note that all the data of Fig.~\ref{fig-3} lie well inside the quasi-1D
regime where $L\gamma_L\ll 1$.

\begin{figure}
  \centering
  \includegraphics[width=0.95\columnwidth]{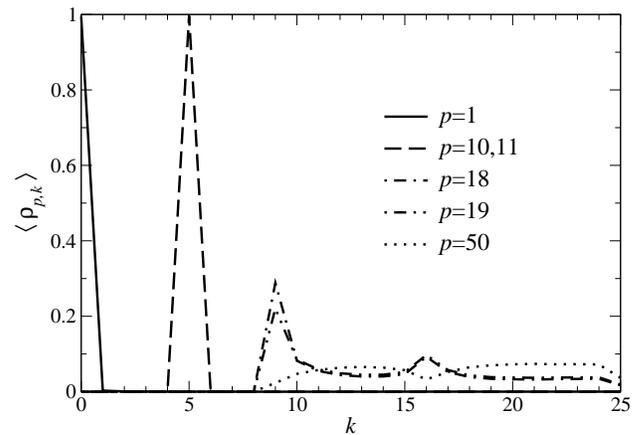}
  \caption{\label{fig-4}
    A sample of numerical results for the average channel weights
    $\langle\rho_{p,k}\rangle$ as function of the channel number $k$ for
    various frame vectors (index $p$).  The strip width is $L=50$,
    energy $E=1.1$ and disorder $w=0.5$. Hyperbolic frame vectors with
    $p\leq 17$ have a large $\langle\rho_{p,k}\rangle\sim 1$ for a specific $k$,
    whereas elliptic frame vectors $p\geq 18$ all have nearly a uniform
    distribution on elliptic channel indices $9\leq k \leq 25$. The
    lines serve as guide to the eye.}
\end{figure}

\begin{figure}
  \centering
  \includegraphics[width=0.95\columnwidth]{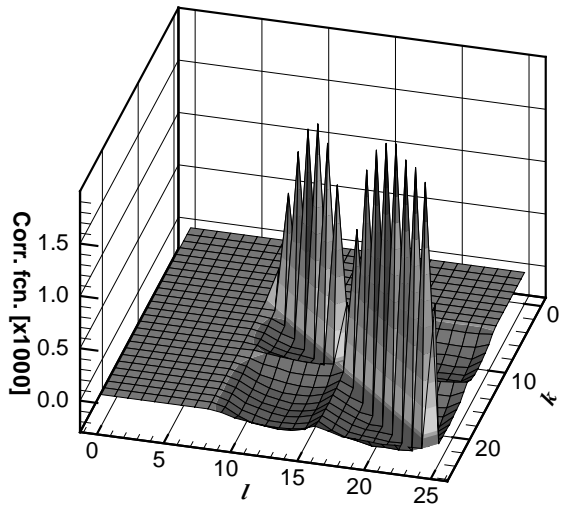}
  \caption{\label{fig-5}
    Contour plot of the numerically evaluated correlation function
    $\langle \rho_{L,l} \rho_{L,k} \rangle - \langle \rho_{L,l} \rangle
    \langle \rho_{L,k} \rangle$ of the $L$th frame vector plotted as
    function of channel indices $l,k$ for energy $E=1.1$, disorder
    $w=0.25$, and strip width $L=50$. The correlation for hyperbolic
    channel numbers $l,k\leq 8$ is zero.}
\end{figure}

We now analyze further quantities of the TMM random dynamical system.
Fig.~\ref{fig-4} shows typical numerical results on the average
channel weights. The first frame vectors align with the hyperbolic
channels. More precisely, they fill them one after another in order of
decreasing dilation exponents. Thus $u_1$ completely aligns with the
expanding direction of the most hyperbolic channel corresponding to the
fundamental of $\Delta_L$. Vectors $p=10, 11$ fill the doubly
degenerate sixth channel. Due to symplectic and orthogonal blocking, the
remaining frame vectors have to be in the elliptic channels. In fact,
they have a more or less uniform distribution over the elliptic
channels (in particular, also the last one $p=50$), only the first elliptic
frame vectors $p=18$, $19$ give more weight to channels
close to the internal band edges. Moreover, further numerics show that all
these distributions are nearly independent of the disorder strength for
$w$ sufficiently small enough. Indeed, one can set up a BGGKY-type
hierarchy for these distributions \cite{Sch03}. The fact that the
distribution over the elliptic channels is nearly uniform is a result of
the mixing properties of the random potential, because the free dynamics
${\bf R}$ gives no preference to any of the elliptic channels.
Fig.~\ref{fig-5} shows that the correlations in the calculation of
Birkhoff averaged channel weights are very small. Hence, as further
approximation, one may factor the stochastic matrix in
(\ref{eq-lyapexpansympgeneral}).

Resuming, the numerical results imply that the distribution of the
last frame vector is uniform on the elliptic channels, the
correlations are weak and all this is uniformly for small $w$.
Hence a good approximation for the coefficient in
(\ref{eq-lyapexpansympgeneral}) should be obtained upon replacing
$\langle\rho_{L,l} \; \rho_{L,k}\rangle$ by $(L_e)^{-2}$ if $L_e$
is the number of elliptic channels.  This gives
\begin{equation}
\label{eq-approx} \gamma_L \;\approx\;
\frac{w^2}{96\,L}\;\frac{1}{L_e^2}\, \sum_{l,k}
\frac{2-\delta_{l,k}}{\sin\eta_l \sin\eta_k} \;,
\end{equation}
where the sum still runs over elliptic channels only. This fits
well with the results of Fig.\ \ref{fig-1} and gives a good
approximation as shown in Fig.\ \ref{fig-2} for $w=0.5$, albeit
not as good as Eq.\ (\ref{eq-lyapexpansympgeneral}). For large
$L$, one may furthermore neglect the $\delta_{l,k}$ in
(\ref{eq-approx}) and approximate the discrete sum by a Riemann
integral. Hence, we infer for $E>0$ and large $L$ (but not too
large so that $E$ stays away from internal band edges)
\begin{equation}
\label{eq-envelop} {\gamma_L} \approx\frac{w^2}{12 L} \left(
\int_{\eta_E\leq\eta\leq\pi} \frac{d\eta}{\pi-\eta_E} \,
\frac{1}{\sqrt{4-(2\cos\eta+E)^2}} \right)^2 ,
\end{equation}
where we set $\eta_E=\arccos(1-{E}/{2})$. Indeed this elliptic
integral (of first kind) can be evaluated numerically. Of course,
this does not give the rich oscillatory structure for finite $L$
anymore as demonstrated in Fig.~\ref{fig-2} for the case $w=0.2$.


We acknowledge financial support by the DFG via SFB 288 and
the priority research program ``Quanten-Hall Systeme''.


\end{document}